# Mixtures of Ethane with $CO_2$ and Water Simulated in ZSM-22: The Role of Polarity and Hydrogen Bonding

**Mohammed Musthafa Kummali** [1, *], **David Cole** [2, a)] **and Siddharth Gautam** [2, b)]

1. Department of Physics, The New College, 147 Peter's Road, Chennai – 600014, India.
2. School of Earth Sciences, The Ohio State University, 125 South Oval Mall, Columbus, OH 43210, USA.
   a) cole.618@osu.edu, b) gautam.25@osu.edu
   \* Correspondence: musthafakummali@thenewcollege.edu.in

**Abstract:** Deciphering the interplay between confinement effects and intermolecular interactions in zeolites is crucial for understanding diverse diffusion behaviors of confined molecules. Recent studies explored the impact of water and $CO_2$ on hydrocarbon dynamics in nanoporous materials. However, differing nanoporous materials, as used in these studies complicate the comparative analysis of $CO_2$ and water effects on hydrocarbons, necessitating a comprehensive investigation with identical confining media and consistent pore diameters. In this study, we investigate the diffusion of ethane, $CO_2$, and water in ZSM-22 molecular sieves. Additionally, we examine the effect of hydration and $CO_2$ on ethane diffusion through the study of ethane-water and ethane-$CO_2$ mixtures. Results indicate enhanced translational motions of $CO_2$ in mixtures, while $CO_2$ mini-mally affects ethane diffusion. In contrast, water is found to slow down the diffusion of ethane by making molecular bridges across the pores. Ethane hampers the translational dynamics of water molecules. Hydrogen bonding in water and the molecular polarity of the fluids are found to play an important role in determining the effects of the presence of one species on the motion of the other. Rotation of the fluid molecules in ZSM-22 is found to occur at two-time scales in both pure state as well as in fluid mixtures. While the short-time fast rotation is determined by the moment of inertia, the long-time rotation is affected by the interaction between fluid molecules and the zeolite atoms. In the case of water, hydrogen bonding hinders rotation and inhibits complete rotation.

## 1  Introduction

Ethane plays a pivotal role in the global energy and petrochemical industries, particularly in the production of ethylene and plastics. As a major component of natural gas, often coexisting with methane in subsurface reservoirs, ethane extraction encounters challenges related to impurities such as $CO_2$ and waters[1–3]. Selective removal of impurities is crucial in extracting high purity ethane for applications in ethylene and chemical derivative manufacturing[4]. Membrane separation, utilizing selectively permeable membranes is a critical technique in gas mixture segregation[5,6]. Engineered nanoporous zeolites are widely recognized for selective adsorption of various types of molecules, such as hydrocarbons, $CO_2$, water etc. In addition to gas separation processes, the selective adsorption of molecules in zeolites including hydrocarbons is useful in catalytic processes, such as the conversion of methane to higher hydrocarbons or the cracking of heavy hydrocarbons to lighter ones[7–9].

Fine-tuning the membrane structure, including pore size and morphology, is crucial for achieving selective separation[10]. This involves understanding the interactions between the membrane material and the adsorbate molecules. Transport properties of the guest molecules through the pores of these nanostructures differ considerably from the bulk behavior[11,12]. Confinement effects resulting from the intricate nanoporous structure of zeolites play a crucial role in altering the transport behavior of the guest molecules[10,13,14]. However, this effect is not solely attributable to geometry, as strong intermolecular interactions between zeolite-adsorbate and adsorbate-adsorbate atoms also influence the diffusion of guest molecules[15]. In certain cases, the confinement effect imposed by the nanoporous structure of zeolites, in conjunction with the interplay between intermolecular interactions, induces intriguing diffusive behaviors[16]. The diffusion dynamics of the adsorbed molecules become even more complex when a mixture of molecules is confined in nanopores. In such cases, the intra-and interspecies interactions are also crucial in determining the diffusion[17]. In general, understanding the interplay between confinement effects and intermolecular interactions is pivotal in unraveling the diverse diffusion behaviors observed in zeolites, ultimately laying the foundation for their effective utilization in a myriad of applications, including gas separation processes[18].

In recent years, several experimental and computational studies have explored the effects of water and $CO_2$, on the dynamics of hydrocarbons confined in zeolite frameworks and other nanoporous media[14,19–28]. These efforts focused on fundamental interactions and mechanisms governing adsorption isotherms[21], competitive adsorption phenomena[28], diffusion dynamics, effect of pore sizes[14], the effect of pressure[24], the influence of inter-crystalline spaces[28], and surface interactions at the molecular level[27]. For instance, Gautam et al.[26] quantified the effect of hydration on the dynamics of hydrocarbons confined in the molecular sieves of MCM-41-S with pore diameters of 1.5 nm. They reported that the presence of hydration impedes the diffusion of propane ($C_3H_8$) molecules, with a more pronounced effect observed at elevated water content within the pores. Similarly, Chathoth et al.[19] investigated the effect of $CO_2$ and $N_2$ on the diffusivity of methane confined in carbon aerogel. Their investigations revealed that the presence of $CO_2$ correlates with an augmentation in the diffusivity of methane within the carbon aerogel matrix. Similarly, Salles et al.[20] reported that $CO_2$ enhances methane diffusion through MIL-47 (V); concomitantly, Gautam et al.[29] observed a similar phenomenon, reporting that $CO_2$ serves to enhance the diffusivity of propane within the pores of silica aerogels. Notably, the silica aerogels in this instance exhibited a range of pore diameters spanning from 15 to 25 nm. Overall, although many research groups have examined the influence of $CO_2$ and water on the diffusivity of various hydrocarbons within systems with confined geometries, these studies have predominantly employed distinct nanoporous materials. Consequently, a comparative study of the effect of $CO_2$ and water on the confined dynamics of hydrocarbons remains far less constrained. This requires investigations wherein hydrocarbon molecules and mixtures diffuse through identical confining media, thereby maintaining consistent pore diameters for a more nuanced understanding of their behavior in confined environments.

While membrane separation with zeolites is widely being used for various gas separation processes, extracting ethane specifically from $CO_2$ and water mixtures using zeolites poses challenges owing to the molecular similarities in size and properties of these compounds. ZSM-22 zeolite is used in gas separation applications due to its unique properties, including a uniform pore size and high thermal stability[30]. In this study, we present a comprehensive analysis of the structure and dynamics, including both translational and rotational aspects, of ethane molecules as they diffuse through the molecular sieves of ZSM-22. Additionally, we explore the effect of $CO_2$ and water on the confinement behavior of ethane within the zeolite. In this study we employ molecular dynamics (MD) simulations for a detailed examination of the dynamic behavior of molecules over time.

The article has been organized as follows: The composition details of the simulated systems, along with the specifics of the simulation techniques employed, are meticulously presented in the 'simulation details' section. The results from the structural and dynamic studies are methodically disclosed in the 'results' section. A detailed examination of the results is presented in the 'discussion' section of the article. Finally, the comprehensive analysis concludes with a concise 'conclusion' of the findings.

## 2 Materials and Methods

The crystallographic unit cell of ZSM-22 contains 24 silicon (Si) and 48 oxygen (O) atoms. Classified as having a Theta-1 structure type, ZSM-22 exhibits one-dimensional ten-membered-ring pores along the Cartesian Z-direction. The pores have cylindrical cross-section with channel diameters 4.5 × 5.5 Å$^2$ (see Fig. 1). ZSM-22 has an orthorhombic crystal structure characterized by lattice constants a = 13.86 Å, b = 17.41 Å, and c = 5.04 Å, as reported in references[31,32]. In this work, the initial simulation cell was generated by replicating a single unit cell of ZSM-22 by 3 × 2 × 6 times along crystallographic axis a, b, and c, respectively, using the visualization software VESTA[33]. In all, nine different systems were simulated differing in the species confined in ZSM-22. Three of these had a single species – ethane, $CO_2$ or water confined in pure states, while the rest had a mixture of either $CO_2$ or water with ethane. Further, the mixture of ethane with $CO_2$ or water were simulated at different compositions where ethane was either minority, equimolar or majority component. Table 1 shows all the systems simulated in this study and Figure 1 depicts a schematic of the pore structure of the zeolite and the molecules studied here.

**Table 1.** Systems simulated in the present study. All systems correspond to different adsorbate species confined in ZSM-22.

| | | Number of molecules of adsorbed species | | |
|---|---|---|---|---|
| **System (adsorbed species)** | **Nomenclature** | **Ethane** | **$CO_2$** | **$H_2O$** |
| Pure ethane | PE | 72 | - | - |
| Pure $CO_2$ | PC | - | 72 | - |
| Pure $H_2O$ | PW | - | - | 72 |
| Ethane - $CO_2$ mixture | MEC-minor | 18 | 54 | - |
| | MEC-eq | 36 | 36 | - |
| | MEC-major | 54 | 18 | - |
| Ethane - $H_2O$ mixture | MEW-minor | 18 | - | 54 |
| | MEW-eq | 36 | - | 36 |
| | MEW-major | 54 | - | 18 |

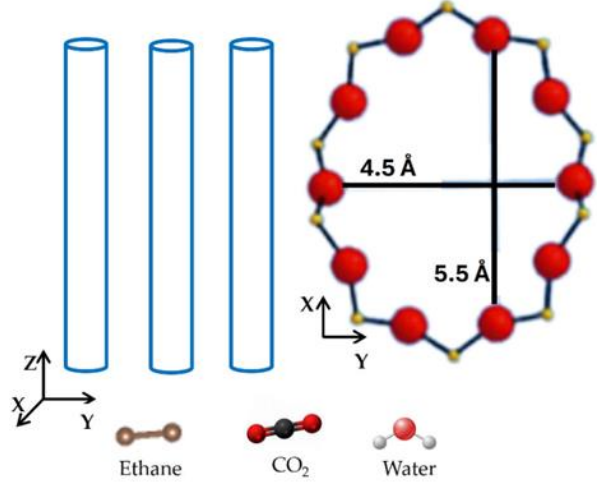

**Figure 1.** Schematics showing the pore structure in ZSM-22 and the molecules studied. ZSM-22 consists of straight isolated channel-like pores with ellipsoidal cross-section oriented along the Cartesian Z-direction. The channel-like pores are formed by a ring structure in X-Y plane formed by oxygen (large red spheres) and silicon (small yellow spheres) atoms of ZSM-22. The adsorbed molecules are ethane (in united atom formalism), $CO_2$ and water.

The interactions between adsorbate molecules and between zeolites and adsorbates were modeled using the Lennard–Jones (LJ) potential, capturing the van der Waals forces and short-range repulsions. Additionally, Coulombic interactions, accounting for electro-static forces, were incorporated between entities featuring partial electrostatic charges. In general, the intermolecular force field took the following form:

$$U_{ij} = 4\varepsilon_{ij}\left[\left(\frac{\sigma_{ij}}{r_{ij}}\right)^{12} - \left(\frac{\sigma_{ij}}{r_{ij}}\right)^{6}\right] + \frac{q_i q_j}{4\pi\varepsilon_0 r_{ij}} \qquad (1)$$

where $\varepsilon_{ij}$ is the depth of the potential well, $\sigma_{ij}$ is the distance at which the intermolecular potential between the atoms i and j becomes zero, the van der Waals radius, and $r_{ij}$ is the distance between atoms i, and j, $q_i$, and $q_j$ are the charges of the i and j atoms.

In the study, interactions between molecules in $CO_2$ and ethane were characterized using the TraPPE force field[34,35]. The united atom formalism was specifically applied to model ethane molecules, connecting two $CH_3$ pseudo-atoms (see Fig. 1). Water molecules were described using the SPC/E force field[36], and the ZSM-22 framework using the CLAYFF force field[37]. All the adsorbate molecules were treated as rigid molecules, with fixed bond lengths set at l(C-$O_C$) = 0.116 nm (where $O_C$ denotes the oxygen atom in $CO_2$), l($CH_3$-$CH_3$) = 0.154 nm, l($O_W$-$H_W$) = 0.100 nm, and l($H_W$-$H_W$) = 0.1633 nm (where $O_W$ denotes the oxygen atom and $H_W$ the hydrogen atom of water, respectively). Throughout all simulations, the positions of all ZSM-22 atoms were fixed. The force-field parameters applied in this research are detailed in an article published earlier[38]. The use of TraPPE and CLAYFF force-fields have earlier provided good agreement between experiments and simulations[28]. The Lennard–Jones (LJ) parameters for the molecules in all studied samples are detailed in Tables A1-A4 in the appendix, and the partial charges on the atoms of water, $CO_2$ and ZSM-22 are presented in Table A5 in the appendix.

Adsorbate molecules were loaded in ZSM-22 using grand canonical Monte Carlo (GCMC) simulations using DL_Monte[39]. Loadings of 72 molecules of a species in the simulation cell are close to the loadings corresponding to a pressure of roughly 1 bar at 300 K. MD simulations were performed using the DL-POLY-4 molecular dynamics simulation package[40]. All simulations were executed in the NVT ensemble at a temperature of 300 K. The Nose–Hoover thermostat was used to regulate the temperature with a relaxation time of 1 ps. The simulations were run for a total of 2 nanoseconds (ns), and a calculation time step of 1 femtosecond (fs) was used. Following the TraPPE-UA convention, a cut-off distance of 14 Å was used. To ensure the equilibrium of the system, an equilibration time of 0.5 ns was used before initiating the production process. The equilibration was verified by monitoring the evolution of the total energy and temperature of the system, which exhibited stable values within acceptable fluctuation limits after 0.5 ns. During the subsequent production process, lasting 1.5 ns, the instantaneous positions, and velocities of all the atoms/pseudo-atoms were recorded at intervals of every 20 fs.

## 3 Results

### 3.1 Distribution of the guest molecules in the zeolite pores

Figure 2 depicts the spatial distribution of pure-state molecules within a randomly selected pore of ZSM-22. In this figure, we present the logarithm of the instances when guest molecules were found at a given location during the entire production time of 1.5 ns. Distribution of molecules in the X-Y plane is shown on the left, while that in the Y-Z plane is shown on the right. The X-Y planes cut across the ZSM-22 pores while these pores run parallel to the Y-Z plane. Elliptically shaped, elongated regions of non-zero intensity in the X-Y plane mirror the structure of ZSM-22 pores, which are slightly elongated along the Y-direction. Ethane molecules exhibit a preference to be near the pore center, indicated by a single high-intensity region at the center (Figure 2 (a)). In contrast, $CO_2$ demonstrates a preference for two locations close to the pore wall and are roughly symmetrically distributed about the pore axis, denoted by two regions of high intensity (Figure 2 (c)). Compared to ethane and $CO_2$, water molecules occupy larger regions of the pore and occupy regions closer to the pore walls while exhibiting a high intensity region close to the center (Figure 2 (e)). In the X-Z plane, periodically spaced adsorption sites alternating between opposite pore surfaces gives rise to a zig-zag pattern for all three fluid species (Figure 2 (b, d, and f)).

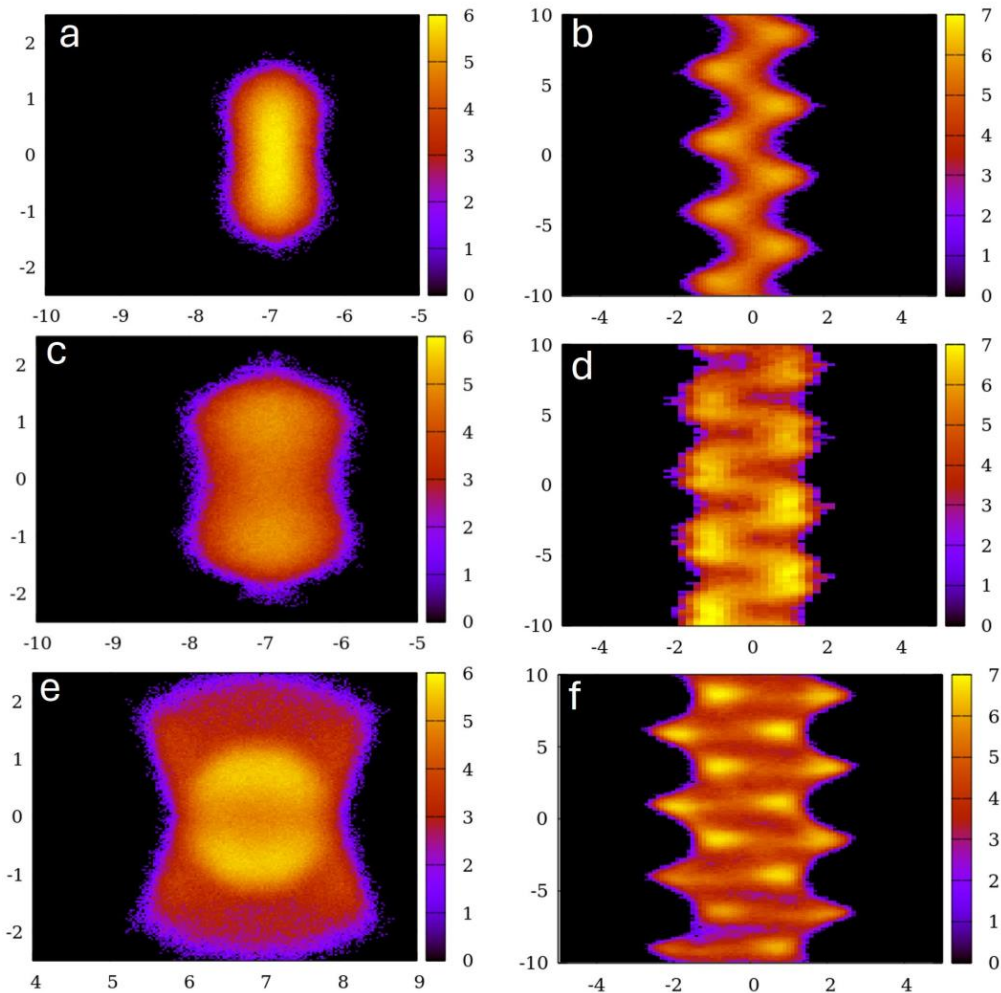

**Figure 2.** The distribution of guest molecules, namely ethane (a,b), $CO_2$ (c,d), and water (e,f), is illustrated in both the X-Y plane (left) and the Y-Z plane (right). The intensity represents the logarithm of the number of times a guest molecule is found at a position with the given coordinates.

## 3.2   Dynamics of the pure-state systems

In Figure 3 (a), the mean squared displacements (MSD) of pure ethane are compared with those of $CO_2$ and water as single components in ZSM-22. The figure illustrates that pure ethane consistently exhibits greater values at all times and demonstrates a more pronounced variation over time compared to the MSD curves of

$CO_2$ and water. The higher MSD values in ethane is expected, given the absence of electrostatic interactions. The translational self-diffusion coefficient (Ds) can be determined using the Einstein relation by fitting the diffusive region of the MSD versus time curve (longer time scale) with a straight line. The steep variation of MSD with respect to time in ethane implies higher values for Ds. Values obtained are: $(7.47 \pm 0.40) \times 10^{-10}$ m²/s for ethane, $(3.92 \pm 0.13) \times 10^{-10}$ m²/s for $CO_2$ and $(6.05 \pm 0.17) \times 10^{-10}$ m²/s for water. Ds values were calculated from the slope of MSD versus time plots in three different time ranges of 100 – 200 ps, 150 – 250 ps and 200 – 300 ps. The values thus obtained were averaged and uncertainty was obtained as the standard deviation over these averages.

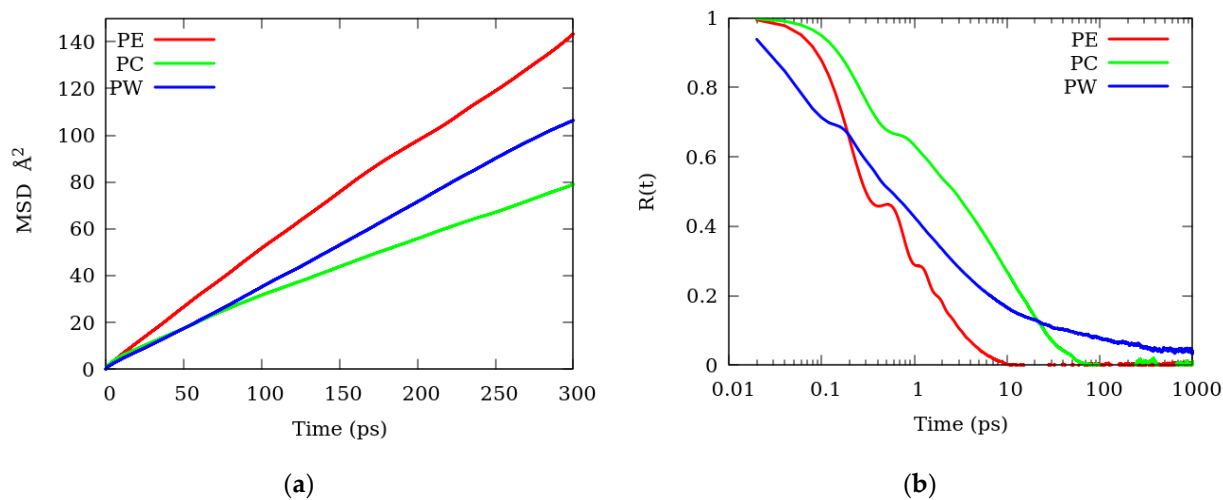

**Figure 3.** MSD is plotted against time for all the pure adsorbate systems under investigation in 3(a) and the corresponding rotational correlation function R(t) in 3(b).

Rotational motion of the guest molecules within zeolite pores were investigated by following the orientation of the molecular axis over time. In particular, the correlation function $R(t) = <u(0) \cdot u(t)>$, where u represents a unit vector attached to the molecular axis, was computed to analyze the rotational behavior over time. R(t) for the three single component systems of ethane, $CO_2$ and water in ZSM-22 are shown in Figure 3(b). For all systems R(t) versus. time curves exhibit two distinct regions separated by a wobble occurring at approximately 1 ps. The short time region corresponds to the initial fast rotation before a typical molecule experiences the presence of any other molecule and is characteristic of the moment of inertia of the molecule. The second region encompassing times above ~ 1 ps corresponds to the long-time overall rotation and it is in this region that the effects of the presence of other molecules can be seen. Figure 3(b) shows a marked difference in behavior of R(t) of the three fluids in the short-time region. The decay rates of R(t) in this region follow the order water (fastest) > ethane > $CO_2$ (slowest). This is because with similar sizes, the moment of inertia in these molecules is predominantly determined by the molecular mass which follow the order water (lightest) > ethane > $CO_2$ (heaviest). In the long-time region above 1 ps, while the temporal evolution of R(t) for pure ethane displays a steep decline, reaching negligible values around 8 ps, the decay in R(t) plot for $CO_2$ is less pronounced, gradually diminishing only at approximately 50 ps. Conversely, in the case of water, the decay of R(t) is more gradual at higher time scales, persisting for several hundreds of picoseconds and exhibiting non-zero values throughout the simulated time. This non-zero value of R(t) indicates the inability of water molecules to completely rotate spanning the entire orientational space during the simulated time. To estimate the time scales of rotational motion, rotational correlation times (τ) are obtained by fitting the long-time region of R(t) with an exponential decay function R(t) = a*exp(−t/τ) + c, where a, τ, and c are fitting parameters. The values of τ thus obtained are: 2.06 ± 0.01 ps for ethane, 13.33 ± 0.02 ps for $CO_2$ and 18.46 ± 0.08 ps for water. The uncertainty is provided by the fitting program.

### 3.3 Effect of the presence of $CO_2$ and water on the dynamics of ethane.

#### 3.3.1 Effect on translational dynamics.

The MSD values of ethane in $CO_2$ mixtures are nearly identical and overlap with the plot for pure ethane, except in the mixture with lower ethane concentration, as indicated in Figure 4(a). In the MEC-minor scenario, the MSD values of ethane are always lower, except at very short times. At these short times, the displacements are ballistic prior to a particle's first collision, and the MSD slope is typically steep. In this regime, the MSD of ethane is the same in all mixtures, including pure ethane.

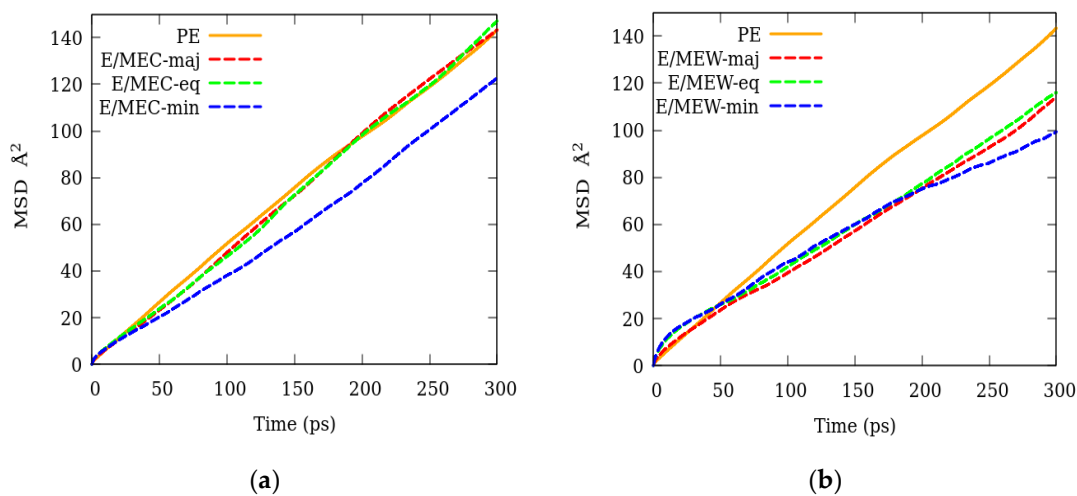

(a)  (b)

**Figure 4.** The MSD plots for ethane in various mixtures, with different concentrations of $CO_2$ (4a) and water (4b), are compared with the MSD of pure ethane.

**Table 2.** The self-diffusion coefficients, expressed in the units of $10^{-10}$ m$^2$/s, are determined from the long-time behavior of the MSD for all the scenarios investigated. E/MEC stands for ethane in MEC mixture and E/MEW for ethane in MEW mixture.

| System | Pure | Major | Eq | Minor |
|---|---|---|---|---|
| PE | 7.47 ± 0.40 | - | - | - |
| E/MEC | - | 8.06 ± 0.67 | 8.20 ± 0.52 | 7.16 ± 0.45 |
| E/MEW | - | 6.08 ± 0.19 | 6.14 ± 0.33 | 4.53 ± 0.67 |

In Figure 4(b), MSD values of ethane in various concentrations of ethane-water mixtures (MEW) are presented. In contrast to ethane-CO$_2$ mixtures, the MSD values in MEW mixtures are smaller than those observed in the pure ethane sample. Furthermore, the temporal variation of MSD is less pronounced for the mixtures when compared to the pure ethane case, particularly at greater times. The self-diffusion coefficient, Ds values obtained for ethane in pure and mixture states are provided in Table 2.

### 3.3.2 Effect on rotational dynamics.

In Figure 5(a) and 5(b), the temporal variations of R(t) for ethane in different concentrations comprising ethane-CO$_2$ 5(a) mixtures and ethane-water 5(b) mixtures are depicted. Additionally, the R(t) plot obtained for pure ethane is included for comparison. In both cases, the R(t) plots of ethane for all the presented samples overlap, suggesting a negligible effect of CO$_2$ and water on the rotational dynamics of ethane. To ascertain the time scales of rotational motion, the second region of R(t) was fitted with the exponential decay function and the resulting rotational correlation times (τ) are listed in Table 3.

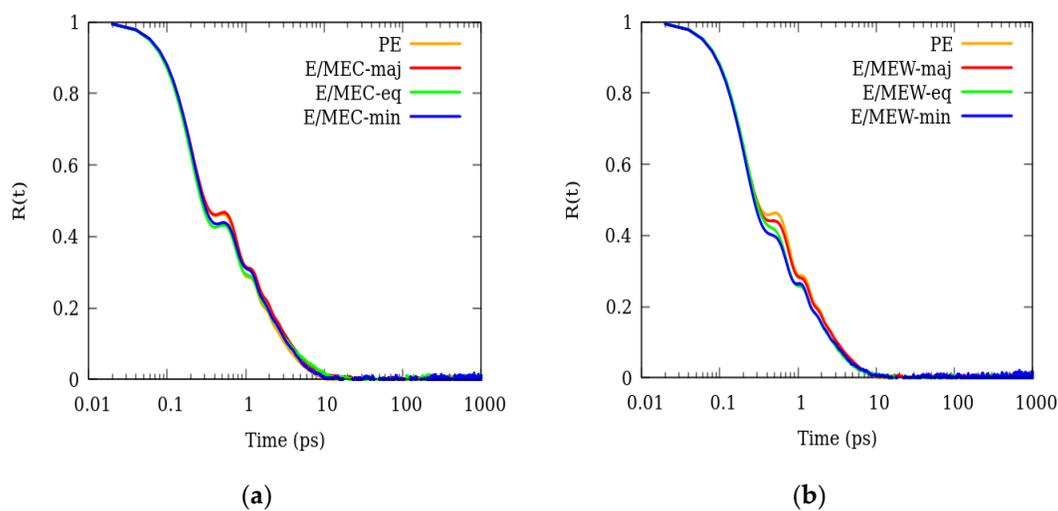

(a)      (b)

**Figure 5.** The rotational correlation function R(t) for ethane in various concentrations of ethane-$CO_2$ 5(a), and ethane-water mixtures 5(b), both plotted against time.

**Table 3.** The rotational correlation times τ (in ps) obtained from the fitting of the rotational correlation functions for ethane in pure and mixture states.

| System | Pure | Major | Eq | Minor |
|---|---|---|---|---|
| PE | 2.06 ± 0.01 | - | - | - |
| E/MEC | - | 2.42 ± 0.01 | 2.25 ± 0.01 | 2.27 ± 0.01 |
| E/MEW | - | 2.18 ± 0.01 | 1.81 ± 0.01 | 1.94 ± 0.01 |

### 3.4 Effect of ethane on $CO_2$

Figure 6(a) illustrates the effect of ethane on the translation of $CO_2$ molecules through the pores. The figure suggests that the presence of ethane enhances the translational motion of $CO_2$ molecules, particularly over extended time scales. However, the influence of varying ethane concentrations on this enhancement is not distinctly evident. The self-diffusion coefficients, obtained by fitting the plots, are detailed in Table 4. The self-diffusion coefficient for pure $CO_2$ in ZSM-22 is $3.92 \pm 0.13 \times 10^{-10}$ m²/s, whereas for $CO_2$ in ethane mixture, values span from $5 \times 10^{-10}$ to $7 \times 10^{-10}$ m²/s. This strongly indicates that the presence of ethane enhances the diffusion of $CO_2$ molecules in the pores. Conversely it is noteworthy that $CO_2$, has a minimal effect on the dynamics of ethane molecules.

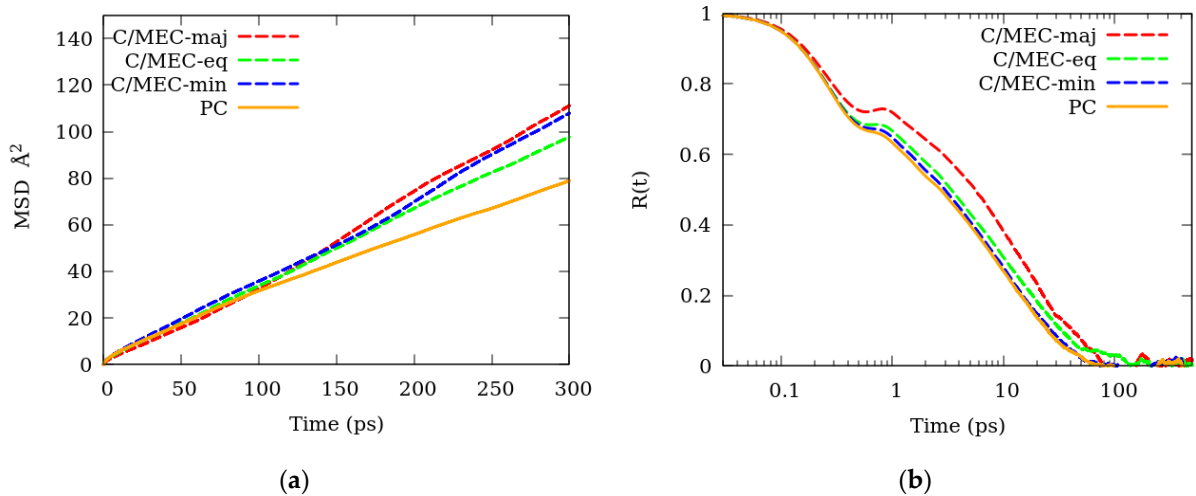

**Figure 6.** MSD (a) and R(t) (b) of $CO_2$ in $CO_2$-Ethane mixture, alongside pure $CO_2$. The notations "maj," "eq," and "min" indicate different Ethane contents in the mixture, with the opposite effect on $CO_2$ content (MEC-maj has lower $CO_2$ content, and MEC-min has higher $CO_2$ content).

In Figure 6(b), the rotational correlation function R(t) for $CO_2$ molecules in the MEC mixture is compared with pure $CO_2$. The figure illustrates that R(t) of pure $CO_2$ molecules decays more rapidly than in mixtures. This observation suggests that the presence of ethane slows down the rotation of $CO_2$ molecules within the pores. The decay constants, extracted from the fitted plots, are also summarized in Table 4. The table indicates higher τ values for the MEC-major and MEC-eq samples with higher ethane contents. This implies that as ethane content increases, rotation dynamics of water molecules slow. This is evident from the flatter curve of the MEC-major sample compared to the steep decay observed in the pure $CO_2$ sample.

**Table 4.** The self-diffusion coefficients, Ds, and the rotational correlation time τ of $CO_2$ in pure and various concentrations with ethane are tabled.

| System / Dynamics | Pure (PC) | Major | Eq | Minor |
|---|---|---|---|---|
| Ds ($10^{-10}$ m²/s) | 3.92 ± 0.13 | 6.58 ± 0.54 | 5.38 ± 0.29 | 6.17 ± 0.59 |
| τ (ps) | 13.33 ± 0.02 | 15.12 ± 0.04 | 15.65 ± 0.05 | 13.77 ± 0.04 |

### 3.5 Effect of ethane on water

Figure 7(a) clearly illustrates that the presence of ethane hinders the translational dynamics of water molecules within the pores. The effect contrasts with the behavior observed for $CO_2$ molecules, where the presence of ethane enhanced translational dynamics. However, the influence of varying ethane concentrations on the inhibiting effect on the translational dynamics of water is not clearly defined. The self-diffusion coefficients, obtained by fitting the MSD plots, are presented in Table 5. Notably, the self-diffusion coefficient (Ds) for pure water is nearly twice than that observed in the MEW mixture. Specifically, the self-diffusion coefficient for pure water in ZSM-22 is approximately $6 \times 10^{-10}$ m²/s, whereas for water in an ethane mixture, the values range from $3 \times 10^{-10}$ to $4 \times 10^{-10}$ m²/s. This implies a discernible reduction in water's translational dynamics in the presence of ethane.

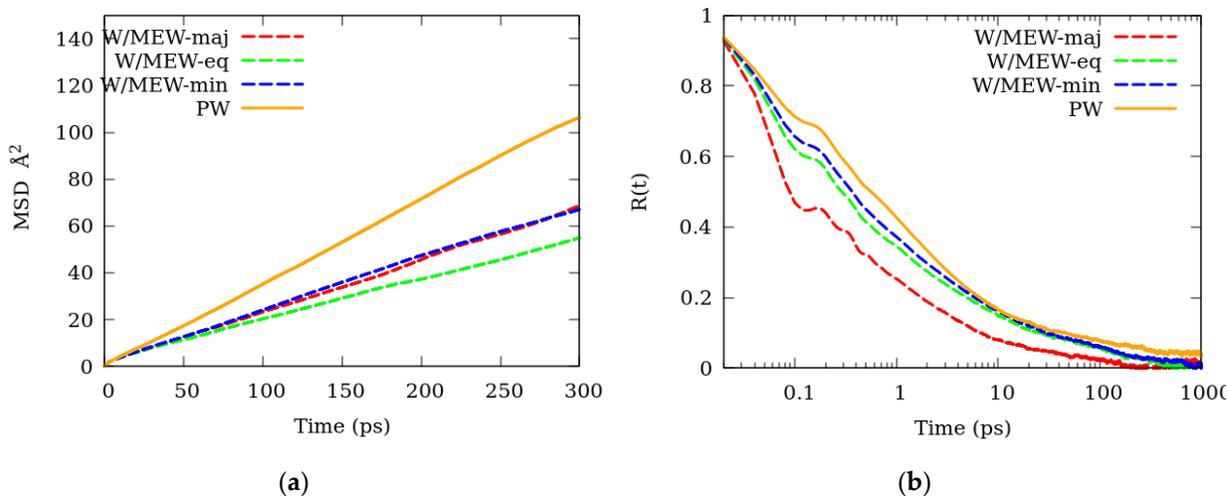

(**a**) (**b**)

**Figure 7.** MSD (a) and R(t) (b) for water molecules at different ethane concentrations are compared. As in figure 6, the labels "maj," "eq," and "min" denote varying concentrations of Ethane in the mixture. Importantly, these labels indicate a converse impact on water content, where MEW-maj corresponds to a lower water content, and MEW-min corresponds to a higher water content.

The temporal variation of the rotational correlation function is presented in Figure 7(b). The figure demonstrates a systematic change as the ethane content of the mixture decreases. The accelerated decay of R(t) persists until around 20 ps, after which the decay curves become parallel until the R(t) of MEW-major sample diminishes at 200 ps. In contrast, R(t) for mixtures with lower ethane content diminishes at longer time scales. This means the presence of ethane enhances the complete rotation of water molecules and this complete rotation occurs at shorter times as the ethane fraction increases.

**Table 5.** The self-diffusion coefficients, Ds, and rotational correlation times τ for water in pure and various concentrations with ethane are tabled.

| Dynamics \ System | Pure (PW) | Major | Eq | Minor |
|---|---|---|---|---|
| Ds ($10^{-10}$ m²/s) | 6.05 ± 0.17 | 3.75 ± 0.19 | 2.84 ± 0.16 | 3.59 ± 0.30 |
| τ (ps) | 18.46 ± 0.08 | 20.17 ± 0.20 | 18.69 ± 0.14 | 17.49 ± 0.06 |

The decay constants, obtained by fitting the plots, are summarized in Table 5. The table indicates a greater τ value for the MEW-major sample with higher ethane content, signifying a slower rotation. This is evident from the flatter curve of the MEW-major sample compared to the steep decay observed in the pure water sample.

## 4 Discussion and Conclusions

Summarizing our results, we show in Figure 8 the diffusion coefficients and rotational correlation times calculated for all the nine systems investigated as functions of ethane fraction.

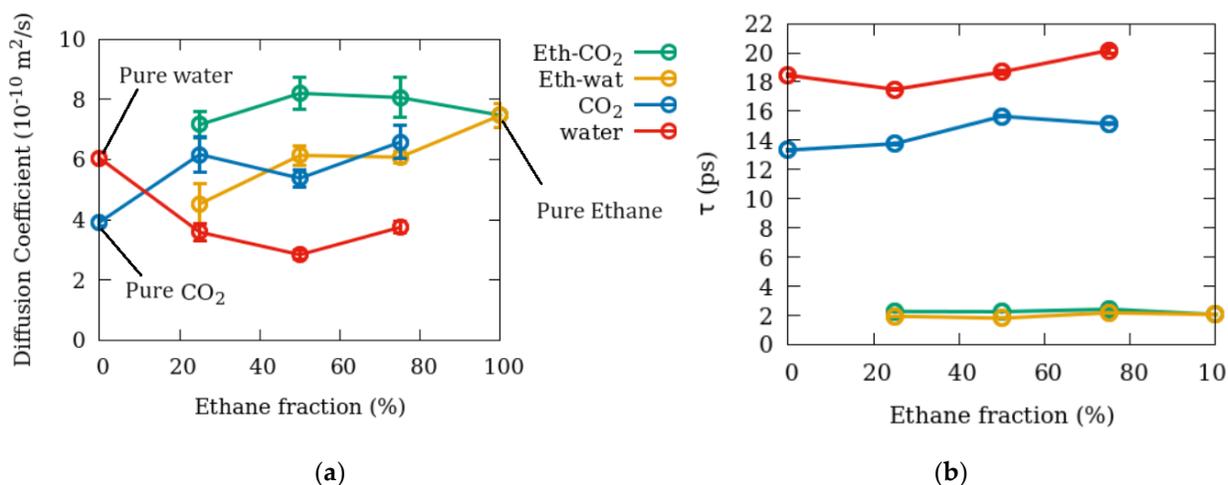

(a) (b)

**Figure 8.** Diffusion coefficient (8(a)) and rotational correlation times τ (8(b)) as functions of ethane fractions.

Diffusion coefficients of the three fluids in pure state confined in ZSM-22 channels as shown in Figure 8 are significantly smaller than those in silicalite – another all-silica zeolite with straight channels of similar dimensions interconnected by sinusoidal pores. For example, at 298 K, using PFG-NMR Bussai et al[41] report a diffusion coefficient of $17 \times 10^{-10}$ m$^2$/s for water confined in silicalite. Further, using MD simulations, they found that while the overall diffusion coefficient of water in silicalite is $33 \times 10^{-10}$ m$^2$/s, it was significantly lower in the X-Z plane, with the slowest component of $7.9 \times 10^{-10}$ m$^2$/s obtained for the Cartesian Z-direction. For ethane in silicalite at 298 K, Bussai et al report a diffusion coefficient of $13 \times 10^{-10}$ m$^2$/s which is higher than that for ethane in ZSM-22 reported here. It is noteworthy that the diffusion coefficients reported here for ethane and CO$_2$ are similar to that in silicalite when the pore connectivity is disrupted by blocking the connecting sinusoidal pores[42]. Thus, connecting the channel-like pores (as in silicalite) leads to higher diffusivities of the confined fluids. It should however be noted that the effects of connecting the pores also depend on the way the pores are connected and the dimensionality of the connecting pores[43].

The effects of ethane fraction on the rotational correlation times are relatively minor. While the decay rates of R(t) in the short-time regime followed the expected trend according to the molecular moment of inertia with water rotation being fastest and $CO_2$ slowest, in the long-time regime the time scales of rotational motion follow a different trend with water being the slowest and ethane fastest. This is because, as stated earlier, in the long-time regime the rotational motion is influenced by the presence of other atoms/molecules. When the interaction between the rotating molecule and other atoms/molecules is strong, the rotation becomes slow. In the present case, water has the strongest interactions with the surrounding zeolite atoms because of its dipolar moment, while ethane-zeolite interaction is the weakest. This explains the difference in the time scales of rotational motion of the three fluids in the pure state.

Ethane, $CO_2$, and water can be seen as representative fluids differing in their electrostatic characteristics. Ethane is non-polar[44], $CO_2$ is quadrupolar[45], whereas water is dipolar[46]. The difference in the electrostatic nature of these fluid molecules leads to a difference in the way they interact with a substrate. While ethane exhibits little or no close interactions with the surface preferring to be in the pore centers, $CO_2$ and water have, respectively, stronger interactions with the pore surface resulting in a more efficient occupation of the available pore volume (see Figure 2). In addition to the dipole moment of water molecules that promote adsorption much closer to the pore walls, the ability to associate with other molecules via hydrogen bonding makes water differ from both ethane and $CO_2$. The intra-species fluid-fluid interactions in the systems studied here suggest a closer packing of water molecules in ZSM-22 channels as compared to ethane or $CO_2$. This is evident from the pair distribution functions (PDF) of the three species shown in Figure 9 (a) that exhibit a peak for the water-water pair at distances shorter than those for ethane-ethane or $CO_2$-$CO_2$ pairs. This closer fluid-fluid interaction between water molecules results in clusters in ZSM-22 channels that are separated by empty space. This contrasts with ethane or $CO_2$ that are distributed relatively more homogenously along the pore axis. This is demonstrated from the simulation snapshots included in Figure 9 (c). As water molecules cluster in the ZSM-22 channels, they form 'molecular bridges' across the pore cross-section thus blocking the passage for any other molecules. This has two consequences – (i) because of the hydrogen bond facilitated cluster formation, the individual water molecules have relatively less freedom to rotate, and (ii) the blockage of the pores due to these molecular bridges impedes the motion of other molecules (e.g. ethane in the ethane-water mixture). This agrees with several other studies reporting the effects of water on the dynamics of confined hydrocarbons[23,26,41].

When ethane is added to water confined in ZSM-22, the tendency of the water molecules to cluster together forming molecular bridges is enhanced. This is evident in Figure 9(b) showing the progressive growth of the first water-water PDF peak on ethane addition at the expense of the second peak. Clustering of water molecules is also evident in the snapshot for the system MEW-Eq shown in Figure 9 (c) where ethane and water molecules seem to be immiscible and occupy two different regions of the pore. Thus, addition of ethane makes water more restricted and thereby less mobile.

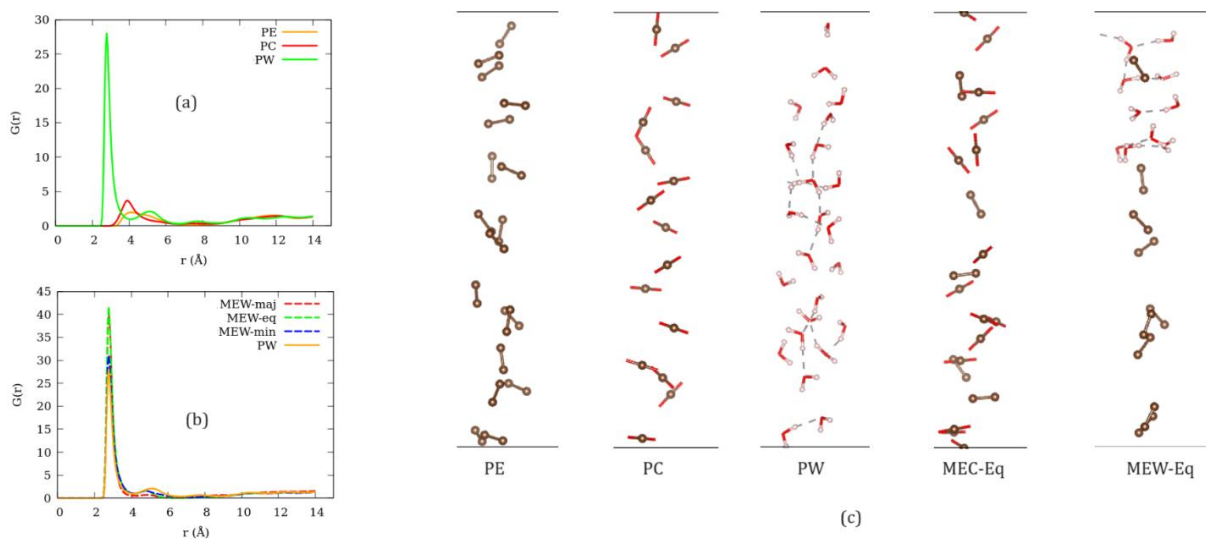

**Figure 9.** (a) pair distribution functions of fluid-fluid pairs in the three single-component (pure) systems. The atoms C(H$_3$), C, and Ow are used as the fluid-centers with only intermolecular pairs used to calculate the functions, ignoring the intramolecular pair C(H$_3$)-C(H$_3$) in the case of ethane (PE). (b) pair distribution function of Ow-Ow pair in the four systems indicated. (c) simulation snapshots of adsorbed molecules in a representative pore from the five systems simulated as indicated. Si and O atoms belonging to ZSM-22 are not shown for clarity.

Unlike water, CO$_2$ molecules do not exhibit strong fluid-fluid interactions within the ZSM-22 channels and tend not to form molecular clusters and hence do not significantly impede the motion of ethane in a mixture (see Figure 9 (c)). This contrasts with other studies that show a significant enhancement of hydrocarbon diffusivity in a mixture with CO$_2$[14,19,20,22,24,25,29]. On addition of CO$_2$, the ethane molecules are displaced from these surface adsorption sites and are relatively free to move, thus explaining the enhancement of hydrocarbon diffusivity on CO$_2$ addition. In ZSM-22 on the other hand, since the pore size is small, the ethane adsorption sites are already close to the pore center because of the interaction with opposite surfaces that nullify each other. This leads to the so-called levitation effect, widely reported in literature[47–49]. Any addition of CO$_2$ therefore does not significantly affect the distribution and hence the dynamics of ethane in ZSM-

22 pores, while CO$_2$ diffusivity is enhanced due to a transfer of kinetic energy from the mobile ethane molecules.

The current study lists the effects of CO$_2$ and water on the dynamics of ethane – a non-polar species confined in narrow cylindrical channels of a siliceous zeolite. Hydrogen bonding is found to play an important role in determining how water affects the dynamics of ethane. In addition to that, the dipole moment of water and the quadrupolar moment of CO$_2$ sets them apart from the non-polar ethane. These properties have been found to play an important role in the competitive adsorption of CO$_2$/CH$_4$/H$_2$O mixtures in brown coal[50] and can be expected to influence the dynamical behavior of adsorbed mixtures too. The results here can thus be applied to other mixtures with components differing in polarity confined in narrow pores.

## 5 Conclusions

MD simulation studies of ethane and its mixtures with water and CO$_2$ confined in straight cylindrical channel-like pores of ZSM-22 are reported here. This study helps us understand the difference in the way the presence of water and CO$_2$ can affect the dynamics of a non-polar species confined in narrow channel-like pores. We find that while the effects of CO$_2$ presence are minimal, the presence of water significantly hampers the motion of ethane. This difference results from the different electrostatic constitution of CO$_2$ and water and the ability of the latter to associate via hydrogen bonding. The efficiency of recovering hydrocarbon fluids from nanoporous silicate rock formations is intricately linked to the interplay between inhibiting and enhancing factors affecting the adsorption and mobility characteristics of various species. The delicate balance between these factors significantly impacts the overall success of hydrocarbon recovery efforts.

**Author Contributions:** Conceptualization, S.G.; methodology, M. M. K.; validation, M. M. K.; formal analysis, M.M.K. and S.G.; investigation, M.M.K. and S.G.; writing—original draft preparation, M.M.K. and S.G.; writing—review and editing, M.M.K., S.G. and D.C.; funding acquisition, D.C. All authors have read and agreed to the published version of the manuscript.

**Funding:** This research was funded by THE U.S. DEPARTMENT OF BASIC ENERGY, OFFICE OF SCIENCE, OFFICE OF BASIC ENERGY SCIENCES, DIVISION OF CHEMICAL SCI-ENCES, GEOSCIENCES AND BIOSCIENCES, GEOSCIENCES PROGRAM, grant number DESC0006878 (D.C. and S.G.).

**Data Availability Statement:** The authors confirm that the data supporting the findings of this study are available within the article.

**Acknowledgments:** We would like to acknowledge STFC's Daresbury Laboratory for providing the package DL-Poly, which was used in this work. Figures in this manuscript were made using the freely available visualization and plotting softwares VESTA[33] and Gnuplot[51].

**Conflicts of Interest:** The authors declare no conflict of interest. The funders had no role in the design of the study; in the collection, analyses, or interpretation of data; in the writing of the manuscript; or in the decision to publish the results.

# 6  Appendix A

**Table A1.** Lennard–Jones parameters for the $CO_2$ –ZSM-22 interactions.

| Interacting Molecules LJ Parameters | Si-C | Si-$O_C$ | O-C | O-$O_C$ | C-C | C-$O_C$ | $O_C$-$O_C$ |
|---|---|---|---|---|---|---|---|
| $\varepsilon_{ij}$ (KJ/mol) | 0.00131 | 0.00225 | 0.38163 | 0.65359 | 0.22400 | 0.38362 | 0.65700 |
| $\sigma_{ij}$ (Å) | 3.051 | 3.176 | 2.983 | 3.108 | 2.800 | 2.925 | 3.050 |

**Table A2.** Lennard–Jones parameters for the ethane –ZSM-22 interactions.

| Interacting a Molecules LJ Parameters | Si-$CH_3$ | O-$CH_3$ | $CH_3$-$CH_3$ |
|---|---|---|---|
| $\varepsilon_{ij}$ (KJ/mol) | 0.00251 | 0.72795 | 0.81500 |
| $\sigma_{ij}$ (Å) | 3.526 | 3.458 | 3.750 |

**Table A3.** Lennard–Jones parameters for the water –ZSM-22 interactions.

| Interacting Molecules LJ Parameters | Si-$O_W$ | O-$O_W$ | $O_W$-$O_W$ |
|---|---|---|---|
| $\varepsilon_{ij}$ (KJ/mol) | 0.00223 | 0.65020 | 0.65020 |
| $\sigma_{ij}$ (Å) | 3.234 | 3.166 | 3.166 |

**Table A4.** Lennard–Jones parameters were additionally used for the mixture –ZSM-22 interactions.

| Interacting Molecules LJ Parameters | C-CH$_3$ | O$_C$-CH$_3$ | CH$_3$-O$_W$ |
|---|---|---|---|
| $\varepsilon_{ij}$ (KJ/mol) | 0.42727 | 0.73175 | 0.72795 |
| $\sigma_{ij}$ (Å) | 3.275 | 3.400 | 3.458 |

**Table A5.** Partial charges on the atoms of water, CO2 and ZSM-22.

| Atoms of Molecules Charge | Si | O | C | O$_C$ | O$_W$ | H$_W$ |
|---|---|---|---|---|---|---|
| q (e) | 2.100 | -1.050 | 0.700 | -0.350 | -0.848 | 0.424 |